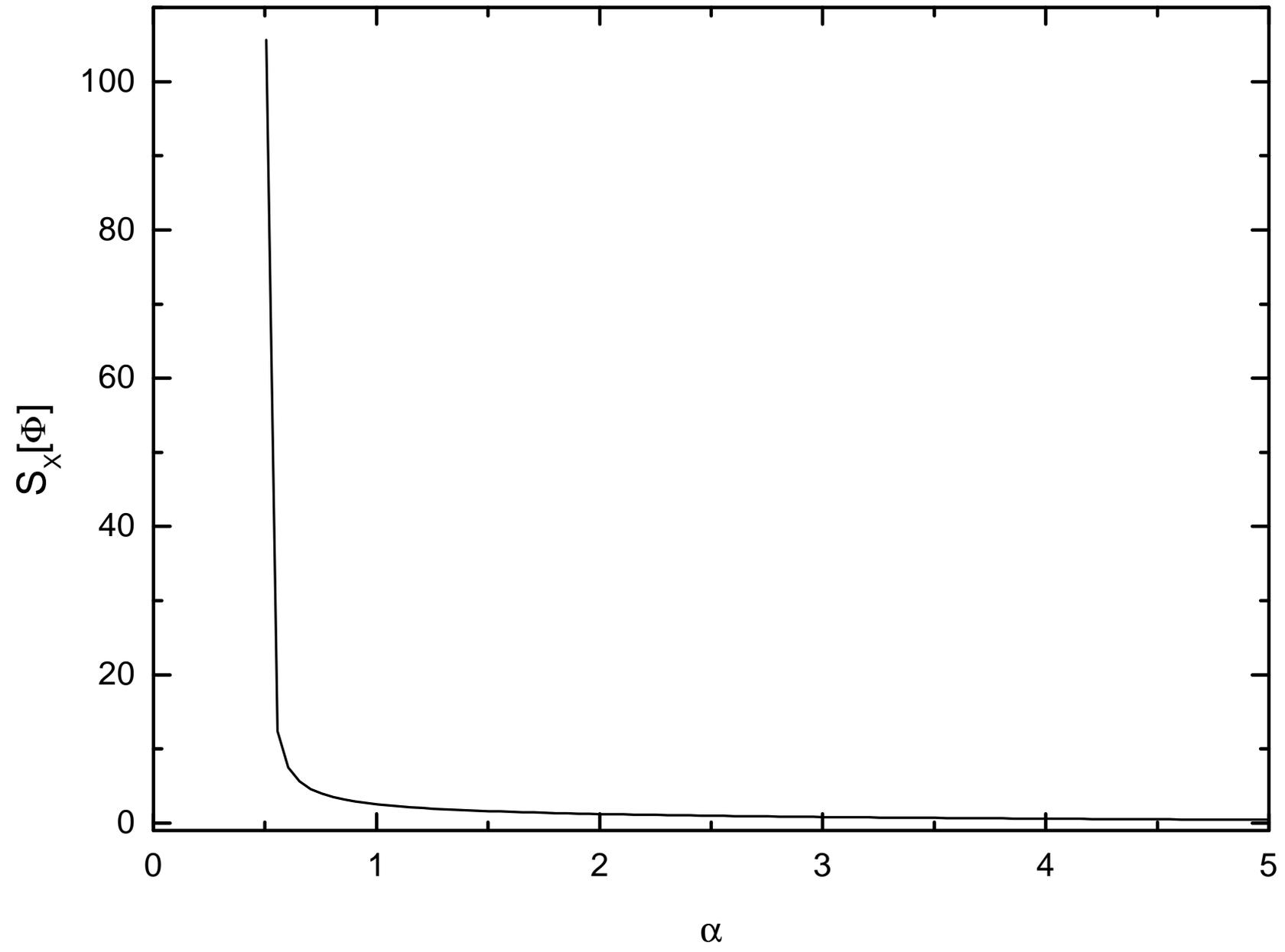

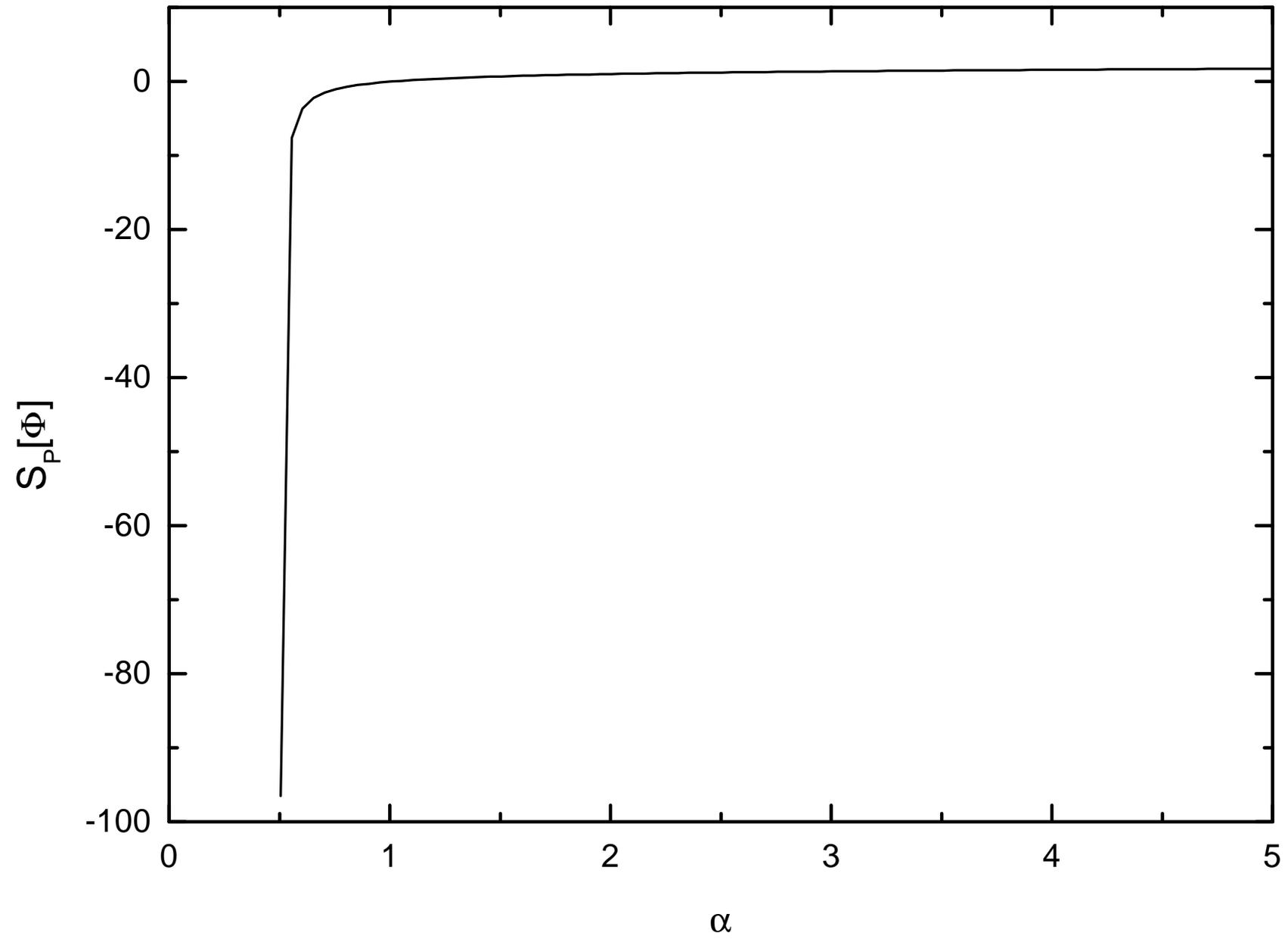

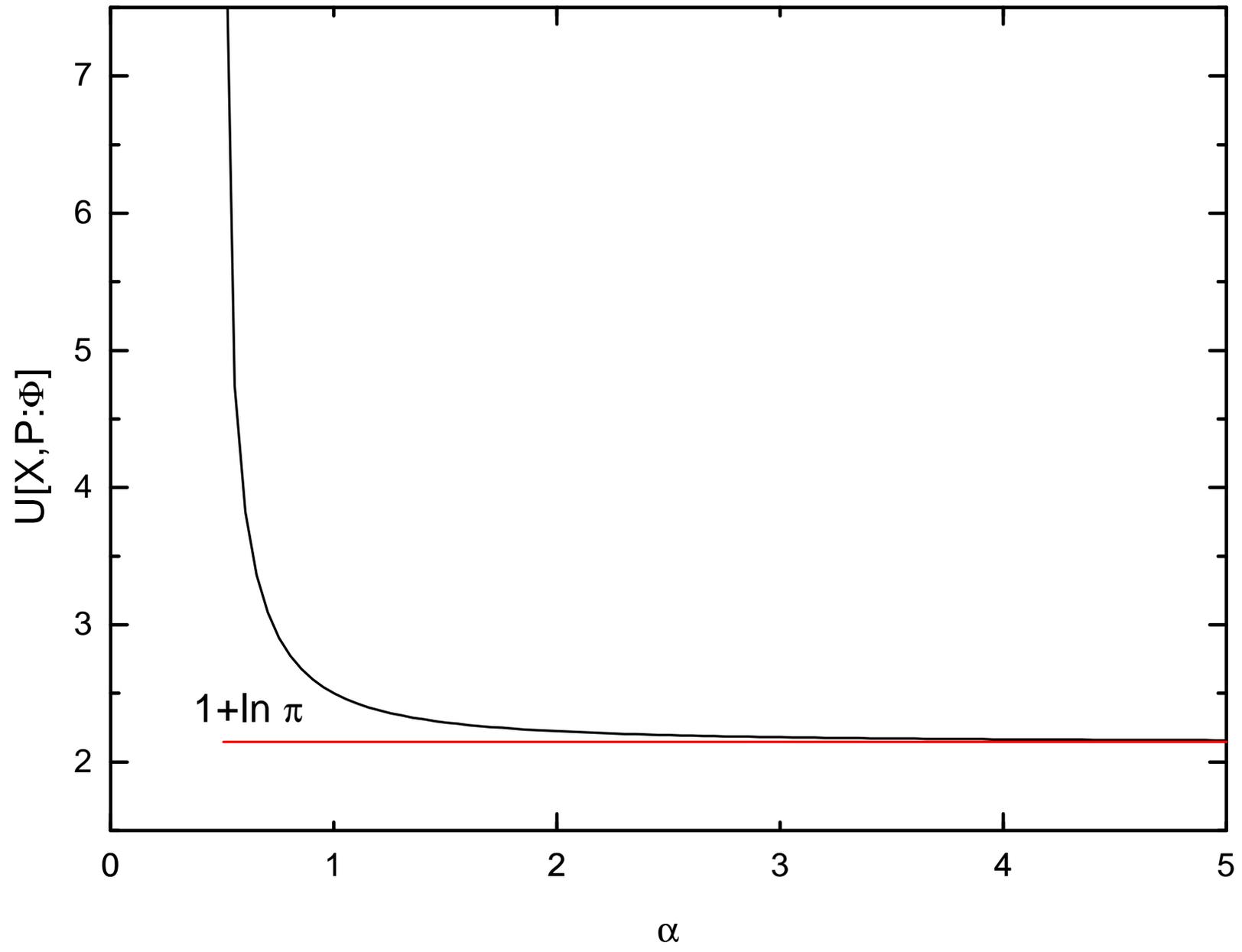

# Entropic uncertainty relation for power-law wave packets

Sumiyoshi Abe[1], S. Martínez[2,3], F. Pennini[2], and A. Plastino[2,3]

[1]*Institute of Physics, University of Tsukuba,*

*Ibaraki 305-8571, Japan*

[2]*Physics Department, National University La Plata, C.C. 727,*

*1900 La Plata, Argentina*

[3]*Argentina National Research Council (CONICET)*

For the power-law quantum wave packet in the configuration space, the variance of the position observable may be divergent. Accordingly, the information-entropic formulation of the uncertainty principle becomes more appropriate than the Heisenberg-type formulation, since it involves only the finite quantities. It is found that the total amount of entropic uncertainty converges to its lower bound in the limit of a large value of the exponent.

PACS: 03.65.-w, 03.65.Bz, 03.67.-a

In ordinary quantum-mechanical situations, the amplitudes of wave packets decrease exponentially at spatial infinities. However, in certain systems, power-law wave packets may be realized. In spite of the fact that they are simple systems, their physical properties have not been fully investigated. In a recent paper [1], Lillo and Mantegna have discussed the free evolution of a power-law wave packet and have found anomalous decay of the maximum of the wave packet, depending on the power-law exponent. (See also Ref. [2].)

To elucidate the physics of the power-law quantum wave packets, it is also important to clarify their statistical properties. In this Letter, we study the uncertainty relation associated with measurements of the position and momentum observables in such a wave packet defined in the configuration space. This is a highly nontrivial issue, since for the power-law wave packet the ordinary measure of uncertainty, i.e. the variance, may be divergent, in general, and therefore the Heisenberg-type formulation of the uncertainty relation is not very useful. To overcome such a difficulty, we employ the information-entropic formulation of the uncertainty relation. We present the analytic expressions for the position and momentum entropies and show that they are finite quantities even if the variance of the position observable diverges. We examine the dependence of the uncertainty on the power-law exponent. We shall see how the total amount of information entropy monotonically approaches the exact lower bound for large values of the exponent.

The power-law quantum wave packet we consider here is given as follows [1]:

$$\phi(x) = \frac{N}{\left(1+x^2\right)^{\alpha/2}}, \tag{1}$$

where $x$ is the dimensionless position coordinate, $\alpha$ the exponent, and $N$ the normalization constant. One sees that this wave function is normalizable in $(-\infty, \infty)$ if

$$\alpha > \frac{1}{2}. \tag{2}$$

Then, the normalization factor is found to be given by

$$N^2 = \frac{\Gamma(\alpha)}{\sqrt{\pi}\,\Gamma(\alpha - 1/2)}, \tag{3}$$

where $\Gamma(s)$ is the Euler gamma function. As noted in Ref. [1], the wave packet in Eq. (1) describes the zero-energy eigenstate of a particle governed by the Schrödinger equation with a potential $U(x)$ of the form

$$U(x) = \frac{\alpha(\alpha+1)x^2 - \alpha}{2\left(1+x^2\right)^2}. \tag{4}$$

Here and hereafter, both $\hbar$ and the mass of the particle are set equal to unity for the sake of simplicity. In the special case when $\alpha = 1$, $\phi(x)$ in Eq. (1) is referred to as the Cauchy wave packet, which has been discussed in Ref. [3].

For $x$ tending to plus or minus infinity, $|\phi(x)|^2$ asymptotically behaves as $|\phi(x)|^2 \approx |x|^{-2\alpha}$. Clearly, the second moment of the position observable diverges:

$$\langle X^2 \rangle_\phi \equiv \langle \phi | X^2 | \phi \rangle = \int_{-\infty}^{\infty} dx \, x^2 |\phi(x)|^2 \to \infty, \tag{5}$$

if the exponent is in the range

$$\frac{1}{2} < \alpha \leq \frac{3}{2}. \tag{6}$$

Therefore, the ordinary Heisenberg-type formulation of the uncertainty relation for the position $X$ and the momentum $P$,

$$\Delta X \cdot \Delta P \geq \frac{1}{2}, \tag{7}$$

is not useful in this range of the exponent. That is, knowing $\Delta X$ yields no information on $\Delta P$, provided that $(\Delta Q)^2 \equiv \langle Q^2 \rangle - \langle Q \rangle^2$ is the variance of the observable $Q$.

For later convenience, we here present the momentum representation (the Fourier transformation) of the wave packet in Eq. (1):

$$\tilde{\phi}(p) = \frac{1}{\sqrt{2\pi}} \int_{-\infty}^{\infty} dx \, e^{ipx} \phi(x) = \frac{2^{1-\alpha/2} N}{\Gamma(\alpha/2)} |p|^{(\alpha-1)/2} K_{(\alpha-1)/2}(|p|), \tag{8}$$

where $K_\nu(z)$ $(\mathrm{Re}\,\nu > -1/2)$ is the modified Bessel function [4].

The information-entropic approach to the uncertainty relation has been repeatedly investigated in the literature [5-13]. A rigorous result relevant to our discussion here is that of Bialynicki-Birula and Mycielski [14]. For measurements of the canonically conjugate pair of the position $X$ and the momentum $P$ of the particle described by the normalized quantum state $|\Phi\rangle$, the associated information entropies are defined by

$$S_X[\Phi] = -\int_{-\infty}^{\infty} dx \, |\Phi(x)|^2 \ln|\Phi(x)|^2, \tag{9}$$

$$S_P[\Phi] = -\int_{-\infty}^{\infty} dp \, |\tilde{\Phi}(p)|^2 \ln|\tilde{\Phi}(p)|^2, \tag{10}$$

respectively, where $\tilde{\Phi}(p) = \langle p|\Phi\rangle$ is the Fourier transformation of $\Phi(x) = \langle x|\Phi\rangle$, as in Eq. (8). Note that these quantities are representation-dependent, in contrast to the von Neumann entropy, $S = -\text{Tr}(\rho \ln \rho)$ with $\rho = |\Phi\rangle\langle\Phi|$. The authors of Ref. [14] have shown that the sum of these quantities has the following optimal lower bound:

$$U[X, P: \Phi] \equiv S_X[\Phi] + S_P[\Phi] \geq 1 + \ln \pi. \tag{11}$$

It is known that this lower bound can be achieved by the coherent state, which also saturates the inequality in Eq. (7). We mention that the uncertainty relation in Eq. (11) has been generalized to the case of finite temperature in Ref. [15]. It is also known [14] that, for any quantum state with the finite variances, the Heisenberg-type relation in Eq. (7) can be derived from Eq. (11).

The superiority of the formulation in Eq. (11) over that in Eq. (7) lies in the fact that both $S_X[\phi]$ and $S_P[\phi]$ always remain finite for the power-law wave packet, $|\Phi\rangle = |\phi\rangle$, in the whole range of the exponent in Eq. (2).

Using Eqs. (1) and (3), $S_X[\phi]$ is analytically calculated as follows:

$$S_X[\phi] = \ln\left[\frac{\sqrt{\pi}\,\Gamma(\alpha-1/2)}{\Gamma(\alpha)}\right] + \alpha[\psi(\alpha) - \psi(\alpha-1/2)], \tag{12}$$

where $\psi(s) = d\ln\Gamma(s)/ds$ is the digamma function [4]. In contrast to the variance $(\Delta_\phi X)^2$, this quantity is in fact finite. Likewise, $S_P[\phi]$ is calculated to be

$$S_P[\phi] = \ln\left[2^{\alpha-2}\sqrt{\pi}\,\frac{[\Gamma(\alpha/2)]^2 \Gamma(\alpha-1/2)}{\Gamma(\alpha)}\right]$$

$$-\frac{2^{3-\alpha}}{\sqrt{\pi}}\frac{\Gamma(\alpha)}{[\Gamma(\alpha/2)]^2 \Gamma(\alpha-1/2)} I(\alpha), \tag{13}$$

where

$$I(\alpha) = \int_0^\infty dp\, p^{\alpha-1} \left|K_{(\alpha-1)/2}(p)\right|^2 \ln\left[p^{\alpha-1}\left|K_{(\alpha-1)/2}(p)\right|^2\right]. \tag{14}$$

Now, from Eqs. (12) and (13), we obtain

$$U[X, P:\phi] = \ln\pi + (\alpha - 2)\ln 2 - 2\ln\frac{\Gamma(\alpha)}{\Gamma(\alpha/2)\,\Gamma(\alpha-1/2)}$$

$$+\alpha[\psi(\alpha) - \psi(\alpha-1/2)]$$

$$-\frac{2^{3-\alpha}}{\sqrt{\pi}}\frac{\Gamma(\alpha)}{[\Gamma(\alpha/2)]^2 \Gamma(\alpha-1/2)} I(\alpha). \tag{15}$$

In Figs. 1-3, we present plots of $S_X[\phi]$, $S_P[\phi]$, and $U[X, P:\phi] = S_X[\phi] + S_P[\phi]$, respectively. Fig. 1 shows that $S_X[\phi]$ monotonically decreases with respect to the exponent $\alpha$, as expected, since a larger value of $\alpha$ makes the wave packet be more localized, leading to a smaller value of the entropy. Correspondingly, Fig. 2 exhibits monotonically increasing behavior of $S_P[\phi]$. An important result is depicted in Fig. 3. There, one clearly appreciates the fact that $U[X,P:\phi]$ monotonically approaches the exact lower bound $1+\ln\pi$ with respect to $\alpha$. It is generally believed that the lower bound is realized only for the Gaussian wave packets. The present result offers another example which asymptotically saturates the information-entropic uncertainty relation.

In conclusion, we have shown that, for the power-law quantum wave packets, the information-entropic uncertainty relation is more useful than the Heisenberg-type relation. We have clarified how the information-entropic uncertainty approaches its lower bound with respect to the value of the exponent of the power-law wave packet.

In the present work, we have employed the Shannon entropy as a measure of uncertainty. It is known that, for characterizing power-law distributions, the Tsallis entropy [16] is a more natural measure than the Shannon entropy. In Ref. [17], the Sobolev inequality, from which the uncertainty relation in Eq. (11) is derived, is discussed in connection with the Tsallis entropy. Moreover, in Refs. [18,19], discussions are developed about the possibility of utilizing the Tsallis entropy as a measure of quantum uncertainty. Investigations in this direction are expected to contribute to a deeper understanding of the uncertainty principle and to further reveal physical properties power-law quantum wave packets.

*Note added*   In a recent paper [20], the information-entropic uncertainty relation as well as the mutual information entropy have been discussed for the multivariate Cauchy-Lorentz distributions and the associated wave packets. There, some interesting scaling behaviors in terms of the number of the degrees of freedom have been found.

S. A. was supported by the Grant-in-Aid for Scientific Research of Japan Society for the Promotion of Science. S. M., F. P. and A. P. gratefully thank the support of CONICET of Argentina. F. P. acknowledges financial support from UNLP, Argentina.

Figure captions

Fig. 1  Plot of $S_X[\phi]$ with respect to $\alpha$ in arbitrary units.

Fig. 2  Plot of $S_P[\phi]$ with respect to $\alpha$ in arbitrary units.

Fig. 3  Plot of $U[X, P:\phi] = S_X[\phi] + S_P[\phi]$ with respect to $\alpha$ in arbitrary units. The horizontal line represents the lower bound $1 + \ln\pi$.